\documentstyle[prb,multicol,aps,epsfig]{revtex}

\begin{document}

\title{Tunable local polariton modes in semiconductors\footnote{Submitted to PRB}}

\author{M. Foygel,$^{1,2}$ Alexey Yamilov,$^1$ Lev I. Deych,$^3$ and A. A. Lisyansky$^1$ }

\address{
$^1$ Department of Physics, Queens College of CUNY, Flushing, NY 11367 \\ 
$^2$ Department of Physics, South Dakota School Mines and Technology, Rapid City, SD  57701 \\
$^3$ Department of Physics, Seton Hall University, South Orange, NJ 07079}

\date{\today}

\maketitle

\begin{abstract}
We study the local states within the polariton bandgap that arise due to deep defect centers with strong electron-phonon coupling. Electron transitions involving deep levels may result in alteration of local elastic constants. In this case, substantial reversible transformations of the impurity polariton density of states occur, which include the appearance/disappearance of the polariton impurity band, its shift and/or the modification of its shape. These changes can be induced by thermo- and photo-excitation of the localized electron states or by trapping of injected charge carriers. We develop a simple model, which is applied to the $O_P$ center in $GaP$.  Further possible experimental realizations of the effect are discussed.
\end{abstract}

%\pacs{71.55.-i,71.36+c,78.30.Fs,71.23.An}
\begin{multicols}{2}

\section{Introduction}

Capture of non-equilibrium charge carriers by a deep defect center provides
an important channel of energy dissipation in wide-bandgap semiconductors
and insulators.\cite{DeepCenters} A significant amount of energy, at least
equal to the binding energy, $\epsilon _{T}\simeq 1eV$, of the electron (or
the hole) to the center, should be released in each capture event, is
usually accompanied by a substantial lattice relaxation. Several mechanisms
can be responsible for the electron transitions involving deep levels in
semiconductors. The energy lost by the captured carrier can be transferred
either to photon(s) in the radiative transition,\cite{Garlick} or to nearby carrier in the Auger effect,\cite{Landsberg} or to a series of long-wavelength acoustic phonons when the carrier descends a staircase of
the excited states in the cascade mechanism,\cite{Lax} or to the local
vibration quanta when multi-phonon emission\cite{HuangRhys} takes place.
There are some indirect\cite{HenryLang,Abakumov,Dean} and direct\cite
{Alt,Skowronski,Lavrov,Lavrov1,GaN,GaAs_Si,GaAs_As,GaAs_Si1,Si_H,AlAs_Be}
evidence that capture or release of the charge carrier is associated not
only with the lattice relaxation but, more importantly with the
alteration of local vibrational modes (LVM's), i.e. with changes in the
local elastic constants.

The subject of the present paper is local excitations of a different type, 
{\it local polariton modes} (LPM's), that are sensitive to the charge-state
induced changes in local elastic constants. These modes arise in polar
crystals in the vicinity of a polariton resonance, where a strong phonon-light
interaction results in splitting of the longitudinal (LO) and transverse
(TO) optic modes. If the spatial dispersion of the TO modes is negative in
all crystallographic directions, there exists a frequency region where the
density of polaritons states vanishes. Defects introduced in such a lattice
may then lead to LPM's inside the polariton bandgap.

LPM's associated with substitutional defects were introduced in Refs. %
\onlinecite{Classics,Podolsky}. They represent electromagnetic excitations
coupled to phonons or excitons with both components, including the
electromagnetic component, localized in the vicinity of the defect. Although
LVM's also interact with the external electromagnetic field, this
interaction results mainly in resonance scattering of light and radiative
decay of the states. Contrary to LVM's, LPM's arise in the polariton
bandgap, where electromagnetic waves cannot propagate. Therefore, there is
neither defect-induced scattering of light nor radiative damping of the
local states. LPM's lead to new optical effects, and strongly affect the
properties of impure crystals. \cite
{Classics,Podolsky,Rupasov,Deych,YamilovEuro,YamilovPRB,YamilovBand,YamilovMC}

In this paper, we show that electron transitions involving deep centers in
semiconductors may lead to a reversible changes of the frequency of the
existing LPM's. We also discuss an even more interesting possibility of
creating/eliminating LPM's by changing the charge-state of a deep center.
When the concentration of these centers is sufficiently high, LPM's develop
into an impurity polariton band (IPB).\cite{YamilovBand,YamilovMC} We show
that in this case the alteration of the local elastic constants can lead to
the creation of an IPB or to the shift of the existing band (and/or to
alteration of its shape). We review materials where these effects may be
observed experimentally.

\section{Local polariton states}

The system under consideration is a polar 3D crystal where dynamics of the
atoms can be described by the classical Newton equations. Polaritons in the
system arise as collective excitations of the polarization waves related to
optical phonons of ``right'' symmetry, coupled to the electromagnetic field
by means of a coupling parameter $\alpha $ proportional to the oscillator
strength of the respective oscillations. The electromagnetic subsystem is
described by Maxwell equations that include the polarization density
related to phonons.\cite{Podolsky}

In a perfect crystal, the solution of the system of the Maxwell and atomic
equations in the long-wave approximation yields the dispersion equation. The
dispersion curves, of course, depend on the symmetry of the crystal. For our
consideration, however, the particular form of the dispersion is not
important as long as the polariton gap exists. Therefore, in the long-wave
approximation we can present the upper, $\Omega _{+}(k)$, and lower, $\Omega
_{-}(k)$, polariton branches in the following isotropic form \cite{Podolsky}
(see Fig. 1): 
\begin{eqnarray}
\Omega _{\pm }^{2}(k)&=&\frac{1}{4}\left( \sqrt{\left[ \Omega _{\bot
}(k)+ck\right] ^{2}+d^{2}}\right. \nonumber \\
&\pm &\left. \sqrt{\left[ \Omega _{\bot }(k)-ck\right]
^{2}+d^{2}}\right) ^{2}.  \label{PolaritonDispersion}
\end{eqnarray}
Here $\Omega _{\bot }^{2}(k)$ is the TO branch of the phonon spectrum, which
in the long-wave limit can be approximated as $\Omega _{0}^{2}-v_{\bot
}^{2}k^{2}$, where $v_{\bot }$ determines the spatial dispersion of TO
phonons, $c$ is the speed of light in the crystal, and $k$ is the wave
vector. The width of the polariton bandgap, $(\Omega _{0},\sqrt{\Omega
_{0}^{2}+d^{2}})$, is determined by a parameter $d$, related to the coupling
constant $\alpha $.
\begin{figure}
\centering
\vspace{-0.1in}
\epsfxsize=3in \epsfbox{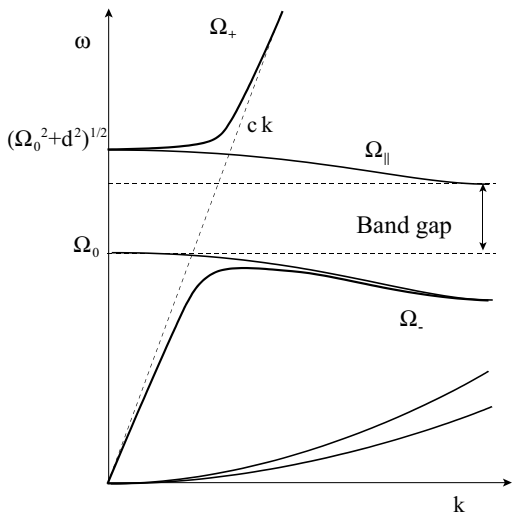}
\vspace{-1in}
\caption{Schematic phonon dispersion curves in a polar crystal.}
\end{figure}

In ideal crystals, LO phonons do not interact with transverse excitations.
In the presence of defects, however, restrictions due to momentum
conservation are relaxed, and the energy of LPM can leak via LO phonons, if
the latter have a nonzero density of states at the frequency of LPM. In this
case, the phonon component of LPM becomes delocalized, but its
electromagnetic component remains localized. There are few crystals where
the dispersion of the LO branch, $\Omega _{||}^{2}(k)=\sqrt{\Omega
_{0}^{2}+d^{2}}-v_{||}^{2}k^{2}$, is not large enough to fill the entire
bandgap and LPM can exist as truly localized states. In most cases, the LO
modes have rather large dispersion with a non-zero density of the phonon
states throughout the entire gap. However, the large dispersion leads to a
relatively small density of the LO states, and the lifetimes of LPM's in
certain materials can be large enough for their survival.\cite{YamilovMC}

The equation for the frequency of LPM, $\Omega _{loc},$ in the presence of a
substitutional defect in a two-sublattice crystal was obtained in Ref.%
\onlinecite{Podolsky}. Assuming that the defect replaces an ion in the
negatively charged sublattice, one can write this equation in the following
approximate form 
\begin{equation}
\frac{4}{3\pi }\frac{(\Omega _{0}^{{}}a)^{3}}{v_{\bot }^{2}c}\frac{d}{\sqrt{%
\Omega _{loc}^{2}-\Omega _{0}^{2}}}\left[ \frac{\delta \beta }{\beta }%
-\left( \frac{m_{+}}{M}\right) ^{2}\frac{\delta m}{\mu }\right] =1,
\label{local}
\end{equation}
where $\delta m$ is the deviation of the mass of the defect from that of the
host atoms, $\delta \beta $ is the local change in the elastic constant, $M$
is the total mass of the positive ($m_{+}$) and negative ($m_{-}$) ions, $\mu =m_{+}m_{-}/(m_{+}+m_{-})$ is their reduced mass, and $a$ denotes the lattice constant. This
equation describes the LPM arising in the vicinity of the TO long-wavelength
limiting frequency $\Omega _{0}$. Obviously, the real-value solution, $%
\Omega _{loc}$, of Eq. (\ref{local}) exists when the expression in the
braces is positive.

The effect of LPM's on optical properties of crystals was studied in Refs. %
\onlinecite{Deych,YamilovEuro,YamilovPRB}. The profile of
electromagnetic-wave transmission is shown to have an asymmetric shape (Fano
resonance) where the maximum is followed by a closely spaced zero. The
maximum value of the transmission exponentially depends on the position of
the defect in the crystal and, without absorption, it reaches unity for the
defect placed at the center of the system. The width of the resonance
decreases exponentially with an increase of the size of the system.

In spite of a general understanding that the local states should produce
resonance tunneling, this result still seems surprising because transmission
of light is affected by structural defects with microscopic dimensions
much smaller than the wavelength of light. The physical explanation of this
effect is based on the fact that the local polaritons emerge due to strong
interaction between the electromagnetic field and local phonons. The latter
have macroscopic dimensions comparable with the wavelength of IR light,
thus making coupling with the external electromagnetic waves effective. As a
result, the electromagnetic wave is carried through the sample by the
phonons that tunnel resonantly due to the presence of the local state.

Because of the large spatial size of the local-polariton states, even at a
very low impurity concentration, $\sim 10^{12}$ $cm^{-3}$, they
significantly overlap. As a result, an impurity polariton band (IPB) is
formed inside the polariton band gap.\cite{YamilovBand,YamilovMC} This band
has a number of interesting properties. For instance, the group velocity of
electromagnetic excitations, propagating via such a band, has been found
proportional to the concentration of the impurities, and it can be
significantly smaller than the speed of light in vacuum. Also, for a large range of defect concentrations, the position of the boundaries
of the IPB linearly depends on the frequency $\Omega _{loc}$ of the ``seed''
LPM.\cite{YamilovMC} Therefore, one can expect that the charge-state induced
changes in the local elastic constants of the deep centers, that generate
the IPB, will affect its boundaries in the same way as they affect the
frequency of LPM at smaller concentrations. In the next section, we will
explore this idea in reference to the well studied substitutional oxygen
defect in gallium phosphide.

\section{Charge-state induced changes in local elastic constants: $O_P$
center in gallium phosphide}

A striking alteration of the local elastic constants was established by
Henry and Lang \cite{HenryLang} in their detailed experimental studies of
the charge states of the $O_{P}$ center in $GaP$. This deep donor center has
two bound states, 1 and 2, with one or two bound electrons, correspondingly.
Henry and Lang concluded that a significant decrease in the local lattice
frequency after capture of the first or second electron is needed in order
to consistently explain a variety of experimental data on photoionization
and thermal emission (deep level transient spectroscopy) involving the two
states in question. For state 2, where the second electron is trapped by or
released from the electrically neutral center with a short-range attraction
potential, this effect can be understood in the framework of the so called
``zero-radius potential'' model.\cite{Abakumov,BazZeld,Meshkov} (Such a
model can be justified if the depth of the impurity potential well for the
second electron at the center is small compared to its binding energy, $\epsilon _{T2}$.\cite{Abakumov}) It can be shown\cite{BazZeld,Meshkov} that
for the ``zero-radius potential'' center the adiabatic potential curves $U\left( q\right) $, corresponding to bound and extended
(continuum) electron states, would rather contact than intersect each other
at the point $q_{c}$ (Fig. 2) where the electron binding energy goes to zero.\cite{Abakumov} (Here $q$ represents the configurational coordinate
corresponding to a single mode of local vibrations that is coupled to the
localized carrier(s).) Let us demonstrate that this rather general
requirement accounts for the alteration of LVM of the $O_{P}$ center in $GaP$
introduced {\it ad hoc } in Ref. \onlinecite{HenryLang}.

In the adiabatic harmonic single-mode approximation, the potential energy of
the heavy ion, which itself is an eigenvalue of the light-electron
Hamiltonian, can be presented as 
\begin{equation}
U_{N}\left( q\right) =\frac{\beta_{0}q^{2}}{2}+N\epsilon \left( q\right)
+U_{c}n_{\uparrow }n_{\downarrow }+\left( 2-N\right) E_{c}.  \label{adpot}
\end{equation}
The first term in this equation describes the elastic energy in the absence
of the localized electrons with $\beta_{0}$ being the elastic constant, and 
\begin{equation}
\epsilon \left( q\right) =\epsilon _{0}-\lambda q-\frac{\gamma q^{2}}{2}
\label{elenergy}
\end{equation}
is the localized-electron energy with electron-phonon coupling taken into
account by expanding the electron energy in powers of $q$ about the
equilibrium point in the absence of electrons. $U_{c}$ is the Hubbard
repulsion energy for two electrons localized at the center, and the last term in
Eq. (\ref{adpot}) is the energy of the electron in the conduction band. $n_{\sigma }=0,1$ is the occupation number of a one-electron localized state with a spin $\sigma $, and $N=\sum_{\sigma }n_{\sigma }=0,1,2$ is the number of electrons trapped by the center. (In Eq. (\ref{elenergy}), we have chosen the negative sign at the quadratic term to assure the {\it internal} contact
of terms $U_{1}\left( q\right) $and $U_{2}\left( q\right) $; the sign of $\lambda $ is irrelevant.)

By using the contact condition at the point $q_{c}$, 
\begin{equation}
U_{1}\left( q_{c}\right) =U_{2}\left( q_{c}\right) ,\;\;\;\left( \frac{dU_{1}%
}{dq}\right) _{q_{c}}=\left( \frac{dU_{2}}{dq}\right) _{q_{c}},
\label{contact}
\end{equation}
it is easy to show that $E_{c}-\epsilon _{0}-U_{c}=$ $\lambda ^{2}/2\gamma $
and therefore the binding (thermal ionization) energy of state 2 is 
\begin{equation}
\epsilon _{T2}=U_{1}\left( q_{1}\right) -U_{2}\left( q_{2}\right) =\frac{%
\lambda ^{2}}{2\gamma \left( 1-x\right) \left( 1-2x\right) }.  \label{epsT2}
\end{equation}
Here $q_{N}$ is the equilibrium configuration coordinate of the center with $%
N$ trapped electrons, and $x=\gamma /\beta_{0}$. By the same token, the
optical ionization energy of state 2 is 
\begin{equation}
\epsilon _{opt2}=U_{1}\left( q_{2}\right) -U_{2}\left( q_{2}\right) =\frac{%
\lambda ^{2}}{2\gamma \left( 1-2x\right) ^{2}}.  \label{epsopt2}
\end{equation}
\begin{figure}
\centering
\vspace{-0.3in}
\epsfxsize=2.5in \epsfbox{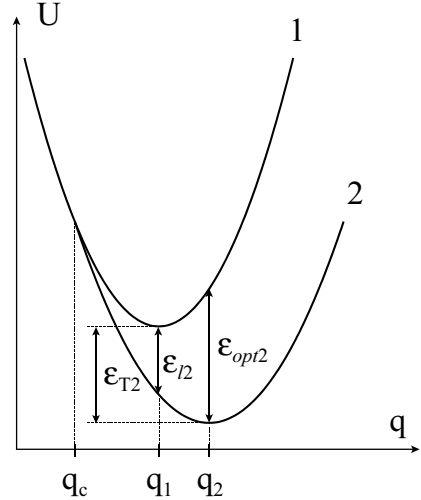}
\vspace{-0.3in}
\caption{Configuration coordinate diagram for the charge states $N=1,2$ of the $O_P$ deep center in $GaP$.}
\end{figure}
Then with experimentally measured values of $\epsilon _{T2}=0.89\;eV$ and $%
\epsilon _{opt2}=2.03\;eV$ for the $O_{P}$ center in $GaP$,\cite{HenryLang}
Eqs. (\ref{epsT2}) and (\ref{epsopt2}) yield $x=\gamma /\beta_{0}=0.36$.
This allows us to immediately evaluate the ratios of the LVM frequencies for
different charge states of the $O_{P}$ center in $GaP$: 
\begin{equation}
\frac{\omega _{2}}{\omega _{1}}=\sqrt{\frac{1-2x}{1-x}}\simeq 0.66;\;\;\;%
\frac{\omega _{1}}{\omega _{0}}=\sqrt{1-x}\simeq 0.80  \label{omegas}
\end{equation}
in fairly good agreement with ratios $\omega _{2}/\omega _{1}=0.65$ and $%
\omega _{1}/\omega _{0}=0.78$ extracted by Henry and Lang \cite{HenryLang}
from numerous experimental data.

The goal of the above simple exercise is to show that if a multi-charge deep
center in one of its states ($N=2$) can be described by the
``zero-radius potential'' model, then non-linear electron-phonon coupling
may result in a substantial change in the local elastic constants associated
with this center when it captures or releases charge carriers. If the second
electron is captured by the neutral center as a result of the radiative
transition, the energy of the emitted photon is 
\begin{equation}
\epsilon _{l2}=U_{1}\left( q_{1}\right) -U_{2}\left( q_{1}\right) =\frac{%
\lambda ^{2}}{2\gamma \left( 1-x\right) ^{2}}.  \label{epsl2}
\end{equation}
From Eqs. (\ref{epsT2}), (\ref{epsopt2}), and (\ref{epsl2}), the following
criterion of applicability of the ``zero-radius potential'' model can be
derived (see also Ref. \onlinecite{Abakumov}): 
\begin{equation}
\epsilon _{T2}=\sqrt{\epsilon _{opt2}\epsilon _{l2}},  \label{crit1}
\end{equation}
as opposed to the standard relation\cite{HuangRhys} 
\begin{equation}
\epsilon _{T2}=\left( \epsilon _{opt2}+\epsilon _{l2}\right) /2,
\label{crit2}
\end{equation}
held for deep centers with linear electron-phonon coupling when no change in
the local frequencies is expected.

If the parameters of a deep center satisfy relation (\ref{crit1}), as it
happens for the $O_{P}$ center in $GaP$,\cite{Abakumov} the capture
(release) of the first or second electron will diminish (increase) the local
elastic constants in the vicinity of this center. To evaluate the
effectiveness of such a rearrangement of the LVM's, let us consider a simple
case of the $p$-type semiconductor doped by shallow acceptors with the
concentration $N_{A}$ and partially compensated by the deep multi-charge
donors of the type considered above with a concentration $N_{D}\ll N_{A}$.
At equilibrium, all the deep donors will be free of electrons (state $0$),
i.e. positively charged. Then, incident light with photon energy close to $E_{g}-\epsilon _{opt1}$ will transfer electrons from the valence
band to the states with $N=0$, thus recharging the deep centers ($%
+\rightarrow 0$). (Here $E_{g}$ is the electron bandgap and $\epsilon _{opt1}
$ is the optical ionization energy of state $1$.) Further evolution of the
system depends on the sign of the effective two-electron correlation energy%
\cite{Anderson,MottDavis} 
\begin{equation}
U_{eff}=U_{0}\left( q_{0}\right) +U_{2}\left( q_{2}\right) -2U_{1}\left(
q_{1}\right) =\epsilon _{T1}-\epsilon _{T2},  \label{Ueff}
\end{equation}
where $\epsilon _{T1}=U_{0}\left( q_{0}\right) -U_{1}\left( q_{1}\right) $
is the thermal ionization energy of state $1.$

By using experimental values \cite{HenryLang} of $\epsilon _{T1}=1.14\;eV$
and $\epsilon _{T2}=0.89\;eV$, it is easy to find that $U_{eff}\simeq 0.26\;eV>0
$ for the $O_{P}$ center in $GaP$. This means that, in this case, the electrically
neutral state $1$ generated by photons with energy ($E_{g}-\epsilon _{opt1}$), which is close to $1.5\;eV$,\cite{HenryLang,Dean} remains metastable under constant illumination conditions and will not be further converted into negatively charged state $2$. Then the electro-neutrality condition, 
\begin{equation}
p+N_{D}^{+}=N_{A}^{-},  \label{neutrality}
\end{equation}
combined with the standard rate equations for the concentration $%
N_{D}^{0}\simeq N_{D}-N_{D}^{+}$ of the deep neutral donors 
\begin{equation}
\partial N_{D}^{0}/\partial t=\sigma _{p1}^{opt}JN_{D}^{+}-p\left\langle
v_{_{{}}p}\right\rangle \sigma _{p1}^{th}N_{D}^{0}  \label{rateD}
\end{equation}
and for the concentration $N_{A}^{0}=N_{A}-N_{A}^{-}$ of shallow neutral
acceptors, 
\begin{equation}
\partial N_{A}^{0}/\partial t=p\left\langle v_{p}\right\rangle \sigma
_{pA}^{th}N_{A}^{-}-e_{pA}N_{A}^{0},  \label{rateA}
\end{equation}
allow one to evaluate the percentage of recharged deep centers. Here $p$ is
the concentration of free holes, $\left\langle v_{p}\right\rangle $ is their
mean thermal speed; $\sigma _{p1}^{th}$ ($\sigma _{pA}^{th}$) is the
non-radiative capture cross-section of the free holes by the deep neutral
donors (shallow negative acceptors); $\sigma _{p1}^{opt}$ is the
cross-section of the optical photo-neutralization of state $0$; $J$ is the
flux of incident photons; $e_{pA}=N_{v}\left\langle v_{p}\right\rangle
\sigma _{pA}^{th}\exp \left( -I_{A}/k_{B}T\right) $ is the rate of thermal
emission of the holes by the shallow acceptors with the ionization energy $%
I_{A}$; $N_{v}$ is the valence-band density of states; $k_{B}$ is the
Boltzmann constant.

From Eq. (\ref{rateA}) it follows that at not very low temperatures such that $T>I_{A}\left[ k_{B}\ln (N_{v}/N_{A})\right] ^{-1}$, all the shallow acceptors are ionized, i.e. the concentration of the free holes [see Eq. (\ref{neutrality})] $p$ is approximately $N_{A}$ ($N_{A}\gg N_{D}$). Then for the steady-state illumination conditions, Eq. (\ref{rateD}) yields 
\begin{equation}
\frac{N_{D}^{0}}{N_{D}}\simeq \left[ 1+\frac{N_{A}\left\langle
v_{p}\right\rangle \sigma _{p1}^{th}}{J\sigma _{p1}^{opt}}\right] ^{-1}.
\label{percentage}
\end{equation}
This means that the deep donors will be almost completely photo-neutralized (%
$+\rightarrow 0$), if the flux of the incident sub-bandgap photons, $J\gtrsim
10^{19}\;cm^{-2}s^{-1}$. (For this estimate, we take $\sigma
_{p1}^{th}=5\times 10^{-21}$ $cm^{-2}$, $\sigma _{p1}^{opt}=1.3\times 10^{-16}$ 
$cm^{-2}$, $N_{A}=10^{17}\;cm^{-3}$, $\left\langle v_{p}\right\rangle =10^{7}$ $%
cm/s$.\cite{HenryLang,Dean}) Such a photon flux can be easily generated by a 
$1W$-source for a spot area of the order of 1 $cm^{2}$. In a $p$-type
semiconductor with positive-$U_{eff}$ centers, this will convert a
high-frequency LVM associated with these centers into a low-frequency one (%
$\omega _{0}\rightarrow \omega _{1}$). However, for a $n$-type material
the photo-neutralization of the deep positive-$U_{eff}$ centers will have
the opposite effect: it converts the low-frequency LVM into the
high-frequency one ($\omega _{2}\rightarrow \omega _{1}$). If the
concentration of the deep centers is high enough ($N_{D}\gg N_{A}$), then
the light from the sub-bandgap or fundamental regions will convert the
intermediate-frequency LVM into the high and low-frequency LVM's: $\omega
_{1}\rightarrow \omega _{0}$, $\omega _{1}\rightarrow \omega _{2}$ [see
Fig. 3a and Eq. (\ref{omegas})].

For materials with negative-$U_{eff}$ centers, the continuous sub-bandgap
(impurity) illumination should, in principle, have more a profound effect on
LVM due to the so called disproportionation,\cite{Alt} i.e.
thermodynamically driven spontaneous decay of the metastable electron state $%
1$ into the states with the next higher, state $2$, and the next lower,
state $0$, number of electrons. In $p$-type materials, it will initially
convert the high-frequency LVM, $\omega _{0}$, into the low-frequency one, $%
\omega _{1}$, associated with the neutral donors, which later on due to $%
U_{eff}<0$ will be further spontaneously converted into the lower one, $%
\omega _{2}$. For $n$-type materials, as in the previous case of positive $%
U_{eff}$, the effect will be opposite: $\omega _{2}\rightarrow \omega
_{1}\rightarrow \omega _{0}$ (Fig. 3b). And, finally, in semi-insulating
material, when the Fermi level is pinned by the negative-$U_{eff}$ defects, 
\cite{Anderson,MottDavis} the illumination should convert the lowest and
highest energy LVM into the intermediate one: $\omega _{2}\rightarrow \omega
_{1}$, $\omega _{0}\rightarrow \omega _{1}$ (Fig. 3b). 
\begin{figure}
\centering
\vspace{-0.1in}
\epsfxsize=3in \epsfbox{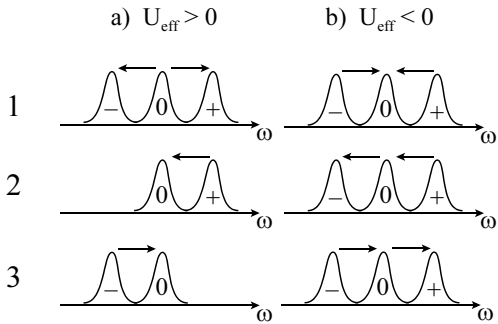}
\vspace{-1.5in}
\caption{Sub-bandgap light-absorption-induced modification of LVM's
for the different charge states of a deep center with (a) positive and (b)
negative effective two-electron correlation energy in the semi-insulating
(1), p-type (2), and n-type (3) semiconductors.}
\end{figure}

\section{Centers and materials perspective for alteration of LPM's}

It is interesting to compare predictions based on our simple model with the
observed optically induced conversion of the charge-state dependent LVM
bands in the oxygen doped $GaAs$.\cite{Alt} It has been proven that the
off-center substitutional $O_{As}$ in $GaAs$ represents one of the few
negative-$U_{eff}$ systems in compound semiconductors.\cite{Alt,Skowronski}
And indeed, during illumination with sub-bandgap light with photon energy around $1.37\;eV$, the high-energy $730.7\;cm^{-1}$ band $A$ is almost
completely converted into the low-energy $714.9\;cm^{-1}$ band $B$ through the
other low-energy $714.2\;cm^{-1}$ band $B^{\prime }$. These bands were
attributed, correspondingly, to the unoccupied, singly, and doubly occupied
electron states of the $O_{As}$ center in certain semi-insulating samples of 
$GaAs$, where the Fermi level was pinned {\it below} the lowest, doubly
occupied, state $2$ of $O_{As}$.\cite{Alt,Skowronski} The material in
question is analogous to $p$-type semiconductors with the negative-$U_{eff}$
centers in our classification. We conclude that, though in accordance
with our model $\omega _{0}>\omega _{2},$ the frequency of the singly
occupied state, $\omega _{1}$, is almost the same as that of the doubly
occupied one, $\omega _{2}$, as opposed to the case of the $O_{P}$ center in 
$GaP$ for which $\omega _{0}^{2}-\omega _{1}^{2}=\omega _{1}^{2}-\omega
_{2}^{2}>0$. An extensive discussion of the electronic structure of $O_{As}$
in $GaAs$ can be found in Ref. \onlinecite{Skowronski}. According to
Skowronski,\cite{Skowronski} oxygen in $GaAs$ creates a complex off-center
defect with dangling gallium hybrids involved, whose coupling strongly
depends on the charge state of the center. The adiabatic potentials of the
coupled dangling bonds in tetrahedrally bonded and amorphous semiconductors
are shown to have a complex multi-well structure that is extremely sensitive
to parameters of the structural defects in question.\cite{PRF,PF,FPA} For
instance, a stretched bond of the type, participating in the formation of $%
O_{As}$ in $GaAs,$ with one or three electrons, will be strongly coupled to
at least two LVM's with charge-state dependent frequencies.\cite{PF} Though
such a defect is not described by the above single-mode model, it can be
responsible for photo-induced changes in the local elastic constants.

Let us now return to $O_{P}$ in $GaP$. The elastic constants for the three
different charge states, $\beta _{N}$ ($N=0,1,2$ enumerates the charge
states of the center), of this center are directly related to the frequency
of the correspondent LVM's given by Eq. (\ref{omegas}): 
\[
\frac{\beta _{2}}{\beta _{1}}=\left( \frac{\omega _{2}}{\omega _{1}}\right)
^{2};\;\;\;\;\;\frac{\beta _{1}}{\beta _{0}}=\left( \frac{\omega _{1}}{%
\omega _{0}}\right) ^{2}.
\]
We want to check if this defect can give rise to LPM's in any of its charge
states. According to Eq. (\ref{local}), the criterion for the appearance of
LPM's is set by 
\begin{equation}
\frac{\delta \beta _{N}}{\beta }-\left( \frac{m_{+}}{M}\right) ^{2}\frac{%
\delta m}{\mu }>0,  \label{LPMtest}
\end{equation}
where $\delta \beta _{N}=\beta _{N}-\beta $ with $\beta $ being the elastic
constant of the host atom. Therefore, it is not sufficient just to know the
ratios of the local elastic constants, we need to establish their values. It
is equivalent to finding the frequencies of LVM's since $\beta _{N}\simeq
m_{O}\omega _{N}^{2}$, where $m_{O}$ is the mass of the oxygen atom. Direct
experimental measurements of these frequencies are not available to us;
however, it is possible to determine them from the configuration coordinate
diagram of this center (Fig. 2). Calculated by means of the ``zero-radius
potential'' model, the multi-phonon emission electron-capture cross-sections
were compared in Ref. \onlinecite{Abakumov} with the experimental data from
deep level transient spectroscopy.\cite{HenryLang} It gave $\omega
_{1}\simeq 195\;cm^{-1}$ for the $O_{P}$ defect in $GaP$. This value differs
from the one ($\omega _{1}\simeq 155\;cm^{-1}$) obtained in the original
paper (Ref. \onlinecite{HenryLang}), where fitting is believed to be somewhat
inconsistent.\cite{Abakumov} Later, in Ref. \onlinecite{Lannoo} it was
argued that because of a strong field dependence of the thermal emission
energy, $\epsilon _{T1}$, of the electron from the singly occupied state,
the deep level transient spectroscopy data should be re-evaluated. This led
to a modified value of the local-phonon energy $\omega _{1}\simeq
290\;cm^{-1}$. We believe that the above re-evaluation of $\omega _{1}$
should not affect the ratios of the frequencies given by Eq. (\ref{omegas}),
for in the framework of the model developed, it is based on the value of
parameter $x=\gamma /\beta_{0}=0.36$. The latter, in turn, was obtained
by means of the ratio $\epsilon _{opt2}/\epsilon _{T2}$ involving
experimentally measured ionization energies of the state $N=2$ with a
short-range potential that can hardly be affected by the electric field in
the area of the p-n junction.

Examining criterion (\ref{LPMtest}) for LPM's to occur, one can see that in
the most favorable case $\delta \beta _{N}$ should be positive while $\delta
m$ should be negative. This has a simple physical explanation. As the frequency of the
TO mode is defined by $\omega_{TO}^{{}}=\left( \beta /\mu \right) ^{1/2}$,
in order to make the defect frequency $\Omega_{loc}^{(N)}$ advance into
the polariton bandgap (above the TO frequency) one should either make $%
\beta_{N}$ larger than $\beta $ ($\simeq 175\;N/m$ in gallium phosphide) of
the host atoms, or decrease the mass of the defect.

For the case of the oxygen center in gallium phosphide where $\delta m<0$,
we examined three values of $\beta _{1}$ obtained by means of all three
values of $\omega _{1}$ from Refs. \onlinecite{HenryLang,Abakumov,Lannoo}.
Then we used ratios (\ref{omegas}) between the rest of the LVM frequencies
to determine the local elastic constants $\beta _{N}$ of the defect in three
charge states. We find that using the data from Refs. %
\onlinecite{HenryLang,Abakumov}, it is not possible to satisfy our condition
Eq. (\ref{LPMtest}). On the other hand, $\omega _{1}\simeq 290\;cm^{-1}$,
obtained in Ref.\onlinecite{Lannoo}, results in $\beta _{0}\simeq
160\;N/m,\;\beta _{1}\simeq 95\;N/m,$ and $\beta _{2}\simeq 40\;N/m$. Even
though for all three charge states $\delta \beta _{N}<0$, $\beta _{N=0}$
still satisfies Eq. (\ref{LPMtest}). Thus, we predict that in the p-type
(semi-insulating) $GaP\!\!:\!\!O$, the LPM associated with $N=0$
charge-state will be eliminated (created) by illuminating the sample with
the light with a photon energy close to $1.48\;eV$ ($0.96\;eV$) (see
Ref. \onlinecite{HenryLang} for details), and as it was shown for the
charge-state dependent LVM's, this process is reversible. We also stress
that the LPM's arise/disappear between TO ($365\;cm^{-1}$) and LO ($405\;cm^{-1}$) frequencies, in contrast to the LVM's that occur either below
or above this region.

The oxygen defect in $GaP$ cannot, by any means, be considered as a single
candidate for tunable LPM's to be observed. Semiconductors and insulators
that possess a complete (omnidirectional) polariton gap include well known
materials such as $GaP,SiC,ZnS,ZnTe,CuI,CaF_{2},SrF_{2},BaF_{2},PbF_{2}$,%
\cite{Bilz,Madelung} as well as extensively studied nitrides $AlN,GaN,InN$.%
\cite{NitridesDisp} In these materials many impurities form deep centers.%
\cite{DeepCenters,Sievers,Weber,Newman} The charge-state dependent LVM's,
which can be considered as precursors of tunable LPM's, remain relatively
unstudied. In a recent paper by Wetzel {\it et al,}\cite{GaN} the
charge-state dependent triplet of LVM's generated by oxygen in $GaN$ has
been reported, which is similar to the $GaP\!\!:\!\!O$ deep center. Even
though this defect does not satisfy our condition for LPM's, Eq. (\ref
{LPMtest}), ($\delta \beta <0$ and $\delta m>0$ for this defect), it gives
us confidence that some defects can give rise to LPM's in crystals with
the complete polariton bandgap. Our optimism is also supported by the fact
that there are many defects that satisfy this criterion but occur in
materials without a polariton gap, namely: $GaAs\!\!:\!\!O$, \cite
{Alt,Skowronski} $EL2$ center in $GaAs$, \cite{GaAs_Si} $GaAs\!\!:\!\!Si$, \cite{GaAs_Si1} $Si\!\!:\!\!H$,\cite{Si_H} and $AlAs\!\!:\!\!Be$,\cite
{AlAs_Be} $Si\!\!:\!\!C$.\cite{Lavrov,Lavrov1} 

To summarize, we have shown that a possibility exists to effectively control
the optical properties, in particular, light transmission, of polar crystals
in the far IR region in the vicinity of the polariton bandgap, by modifying
the charge state of the deep center by means of light from the visible or near IR region. To obtain more specific results, further experimental and theoretical studies are needed.

\section*{Acknowledgments}

We are indebted to S. Schwarz for reading and commenting on the manuscript.
This work was partially supported by the NATO Linkage Grant N974573, CUNY
Collaborative Grant, and PSC-CUNY Research Award, as well as by the NSF
grant DMR-0071823 and the Nelson grant (SDSMT).

\end{multicols}

\end{document}